\documentclass[
]{ceurart}

\usepackage{listings}
\usepackage[T1]{fontenc}
\usepackage{graphicx}
\usepackage{tabularx}
\usepackage{fancyvrb}
\usepackage{fvextra}
\usepackage{float}
\usepackage{tcolorbox}
\usepackage{multicol}
\usepackage{multirow}
\begin{document}


\copyrightyear{2026}
\copyrightclause{Copyright for this paper by its authors.
  Use permitted under Creative Commons License Attribution 4.0
  International (CC BY 4.0).}

\conference{Joint Proceedings of REFSQ-2026 Workshops, Doctoral Symposium, Posters \& Tools Track, and Education and Training Track. Co-located with REFSQ 2026. Poznan, Poland, March 23-26, 2026.}

\title{User Reviews as a Source for Usability Requirements: A Precursor Study on Using Large Language Models}

\author[1]{Cedric Wellhausen}[%
email=cedric.wellhausen@gmail.com
]

\author[1]{Laura Reinhardt}[%
email= laura.reinhardt@inf.uni-hannover.de
]

\author[1]{Kurt Schneider}[%
email= kurt.schneider@inf.uni-hannover.de
]
\address[1]{Leibniz University Hannover, Software Engineering Group, Hannover, Germany}


\begin{abstract}
\textbf{\textbf{[Context and motivation]}} It is known that user-centered approaches to requirements engineering in general lead to a better suited product for the end-users. LLM4RE provides promising approaches to support the requirements elicitation process (e.g. classification of requirements). \textbf{[Question/problem]} Previous approaches focus on Machine-Learning (ML) or Deep-Learning (DL) aspects, which require intensive training with a large amount of manually labeled data. LLMs, on the other hand, are pre-trained on large amounts of user-generated text data, enabling a user-centric workflow to analyze requirements.
\textbf{[Principal ideas/results]} In this paper, we explore the possibility of exploiting the
improved natural language understanding of LLMs, rather than strict ML classification, together with the mass extraction of user reviews to analyze if the performance of LLMs in understanding user reviews is comparable to the performance of human raters. This enables a quick and cheap workflow for development teams to gather and process their user’s requirements.
\textbf{[Contribution]}
This paper provides three major contributions: (1) We provide a completely coded dataset of 300 user reviews containing usability-relevant aspects from three different types of apps, that were labeled by two human raters and by an LLM. (2) We build an initial prompt, based on two prompt engineering iterations and specifically developed coding guidelines derived from the 10 Nielsen Usability Heuristics, for LLMs to filter usability relevant user reviews. (3) We determine that LLMs are generally able to recognize usability as a non-functional requirement in user reviews, in terms of their F-score, but the performance and reliability is strongly dependent on the prompt.

\end{abstract}

\begin{keywords}
    Usability \sep
    LLM \sep
    Requirements Engineering
\end{keywords}

\maketitle

\section{Introduction}

The increasing functional complexity of modern software products presents major usability challenges. Although integrating user-centred approaches into requirements engineering can mitigate these issues and lead to a better suited product for end-users \cite{bakiu}, directly involving users remains a resource-intensive process that is often unfeasible for development teams. To bridge the gap between development teams and users, Crowd-based Requirements Engineering (CrowdRE) leverages publicly available feedback from forums and app stores to indirectly capture user needs from user feedback \cite{crowdre}.

Machine-Learning and Deep-Learning approaches for classifying requirements are frequently combined with CrowdRE or crowd-sourcing methods~\cite{nlp4re,llm4re,unterbusch}. One significant limitation of these approaches in classifying user needs out of user feedback is, that these approaches often rely on large, manually labeled datasets, which creates a significant bottleneck for rapid development cycles \cite{nlp4re,llm4re,unterbusch}. Nowadays, new approaches involving Natural Language Processing (NLP4RE) and Large Language Models (LLM4RE) gain more attention, due to the fact, that LLMs are able to process a large amount of information, without requiring any additional training data for most purposes \cite{finetuning}. Despite the rise of Natural Language Processing (NLP4RE) \cite{nlp4re} and Large Language Models (LLM4RE) \cite{llm4re} in requirements engineering tasks, a critical research gap remains: combining the effectiveness of modern LLM language understanding with the user-centric methods of CrowdRE. 

In this work, we address the potential of LLMs to extract and process user requirements from mass review data without the need for extensive task-specific training. LLMs can be instructed with natural language, where the instructor gives an LLM a so-called \textit{prompt}~\cite{intelligent_agents}. We want to compare if the performance of an specifically prompted LLM is comparable to the performance of human coders to identify usability-relevant user reviews. Furthermore, the LLM will be analyzed in context of their reliability to identify usability-related user reviews with an tailored prompt, specifically developed for these kind of tasks. This work will test the mass elicitation in light of usability requirements which can (i) be very nuanced and (ii) play an important role in product marketing and user adoption.
\section{Background and Related Work}

\subsection{Usability}

Usability by ISO 9241-11 is defined as the extent to which a product, including mobile apps, can be used by specified users to achieve their specified goals~\cite{iso20249241}. This also includes factors such as learnability and satisfactory in use~\cite{iso20249241}. Furthermore, usability can be defined as the general capability of an entity to being used~\cite{usabilitydefinition, QUINONES201789}. Due to the widespread definitions and applicability of usability in all kinds of systems, it is quite difficult to establish a single, universally accepted method, necessitating an iterative and context-dependent approach to evaluation. One of the more established methods to evaluate the usability of software systems is the \textit{10 Usability Heuristics} by Nielsen~\cite{ten_ue_heuristics, nielsen1994enhancing}. These heuristics include a set of ten general principles on how to design user interfaces~\cite{ten_ue_heuristics, nielsen1994enhancing}. Additionally, usability problems in real systems can be derived from these heuristics~\cite{nielsen}. 

\subsection{Prompt Engineering}

Prompt Engineering is an emerging field that aims to produce an input for a Large Language Model (LLM) such that the LLM will produce a desired output when given the input~\cite{white2023promptpatterncatalogenhance}. Such an input is then called a \textit{prompt}. A textual prompt is a set of instructions, given to a LLM, to customize, enhance and refine its capabilities~\cite{white2023promptpatterncatalogenhance, prompt}. This adaptability of prompts as an input for LLMs is inherently different from traditional machine learning, where model retraining is often required for specific tasks~\cite{sahoo2025systematicsurveypromptengineering}. Generating a fitting prompt for a desired output can not be perfectly formalized but is approached with a set of techniques that have been shown to work well in practice (e.g. iterative prompt engineering~\cite{schulhoff2025promptreportsystematicsurvey}, chain-of-thought prompting~\cite{Wei}). One technique, iterative prompting, involves direct human feedback, also known as human-in-the-loop, to create an initial draft of a prompt that is then further refined by the humans~\cite{schulhoff2025promptreportsystematicsurvey}. The chain-of-thought prompting technique aims to improve LLMs' outputs by prompting an LLM to describe its intermediate steps while performing tasks, thereby making its approach transparent for further analysis~\cite{schulhoff2025promptreportsystematicsurvey, Wei}.

\subsection{Related Work}

Bakiu et al.~\cite{bakiu} have developed a Machine-Learning-based method for extracting usability aspects from user reviews. To this end, a classifier was trained with manually classified reviews from the categories \textit{Software} and \textit{Video Games}. To do this, they used four sets of usability dimensions, including the five dimensions of Nielsen~\cite{nielsen_book} as one of the sets. The findings revealed, that only analyzing user reviews can not be replaced by already existing methods. Furthermore, their method to extract usability aspects was provisionally evaluated with the result that further and larger studies are necessary to gain an accurate picture of its effectiveness.

Hedegaard et al.~\cite{hedegaard} conducted a study collecting reviews from the categories \textit{Software} and \textit{Video Games}. They used 2972 reviews from the video game sector and 520 reviews from the software sector and used different models to extract usability aspects. Therefore they used one model (CLASSICUA) based on the usability definition according to Nielsen~\cite{nielsen_book} and another model (FREQUENT) based on terms that are frequently associated with usability. The CLASSICUA classification resulted in usability-relevant aspects being mentioned in more than 40\% of the reviews. FREQUENT achieved a value of up to 30\%. 

Previous research has concentrated exclusively on machine-learning (ML) approaches to the classification of usability-relevant aspects. However, such approaches require a substantial quantity of labeled training data.  In the context of our research, there is a clear objective to transition the perspective from the utilization of resource-intensive machine learning algorithms to the deployment of more accessible Large Language Models (LLMs). While ML is only able to simply classify usability aspects on simple keywords (e.g. ''slow'' or ''error''), LLMs are able to understand contextual information and informal language like emojis (e.g. ''I had to click the Back-Button three times :-/''). This ability facilitates the identification of usability aspects in a more diverse range of written statements. In addition, LLMs have the capacity to justify their decisions, a capability that is absent in the case of ML, which provide solutions without offering any reasoning to support their decisions. Our work is also supposed to analyze different types of software, that are more integrated in daily life. Therefore, we not only included reviews from software in general, but rather specific software used in maps and navigation, music-composition or E-commerce \cite{SoftwareTypes}.

\section{Research Design}
\label{chapter-3}

\subsection{Research Questions}

Our research in the work is framed by the following research questions:
\newline\newline
\noindent
\textbf{RQ1:} Is the performance of an LLM, when prompted with a specifically tailored prompt to identify usability-related user reviews, comparable to the performance of a human coder?
\newline
\noindent
\textbf{RQ2:} What is the reliability of an LLM for identifying usability-related user reviews in a random dataset of user generated app reviews in case of specifically tailored prompts?
\newline
\newline
\noindent
In this paper, we aim to investigate whether the LLM gpt-4.1 by OpenAI can be used as an effective and resource efficient tool to support requirements elicitation in the context of CrowdRE. Our focus lies on the non-functional requirement usability as this allows us to (1) test the LLM in a controlled environment that can be defined in terms of norms and heuristics and (2) to see in what extent the LLM understands user reviews, as these are in general unstructured and extracting usability-related information has proven difficult~\cite{bakiu}.


\subsection{Data Collection}

To test the capabilities of the LLM on real-world data, we collected a dataset of 300 user reviews. The 300 user reviews originate from the Google Play-Store and are evenly distributed over three apps: \textit{BlitzerDE}, \textit{Lidl Plus}, and \textit{FL Studio}.
The three apps were selected based on the criteria of user numbers, ratings and number of ratings. Both criteria were intended to maximize the diversity of individual reviews such that a broad range of opinions could be covered. Furthermore, apps from different software types and specific categories were selected in order to obtain a broad picture across various domains. Table~\ref{tab:app_stats} shows the ratings, user counts and specific software type for all three apps.

\begin{table}[h]
    \centering
    \caption{Overview of the apps which served as sources for user reviews, their respective Google Play-Store analytical data and their type of software.}
    {\renewcommand{\arraystretch}{1.2}
    \begin{tabularx}{\textwidth}{l|l|l|l|X} 
    \hline
        App & Rating & User Count & Ratings Count & Software Type \cite{SoftwareTypes}\\
        \hline\hline
        BlitzerDE & 4.7 & 10 million & >120k & Information display and transaction entry (maps and navigation)\\
        Lidl Plus & 3.8 & 100 million & >1.70 million & Information display and transaction entry (E-commerce) \\
        FL Studio & 4.0 & 1 million & >40k & Computation-dominant (music composition) \\
    \hline
    \end{tabularx}
    }
    \label{tab:app_stats}
\end{table}

\noindent
Besides the aforementioned criteria, we also included the genre and intended target audience in our decision. All three apps are targeted towards adult users but in very different scenarios. While BlitzerDE is used in an automotive-centric scenario, Lidl Plus's users will use the app before and while shopping for e.g. groceries. FL Studio the other hand is an app that is aimed at musicians to help them produce music. As these different scenarios affect how the usability of the app is perceived by the users, we assume the user reviews will contain vastly different opinions and feedback. The goal is to construct a dataset that has a high diversity of user reviews which then can be used to test the LLM with as little bias towards certain kinds of user reviews as possible.

\subsection{Data Analysis}

The data analysis process of this paper is divided into two phases. The first phase serves the purpose of (1) building a fully labeled dataset which can be used to evaluate the performance of an LLM, and (2) establishing a performance baseline that represents the performance of usability-experts and can be used to evaluate the LLM. The second phase uses the results of the first phase to then answer both research questions RQ1 and RQ2.
\newline\newline
\noindent
\textbf{First Phase:} Figure~\ref{fig:dc-process} shows the process that was developed for the first phase. Labeling the dataset began with establishing a set of indicators for usability. For this, we used the \textit{10 Usability Heuristics} according to Nielsen~\cite{ten_ue_heuristics, nielsen1994enhancing} as well as a few indicators we defined as not usability-relevant. These are: (i) Feature-Requests, (ii) overlap with adjacent non-functional requirements such as accessibility and compatibility, and (iii) vaguely formulated statements for which too much interpretation by the raters would haven been necessary. We decided to use the 10 Usability Heuristics as they define usability very detailed when compared to other definitions like the ISO 9241~\cite{iso20249241}.

\begin{figure}[h]
    \centering
    \includegraphics[width=0.9\linewidth]{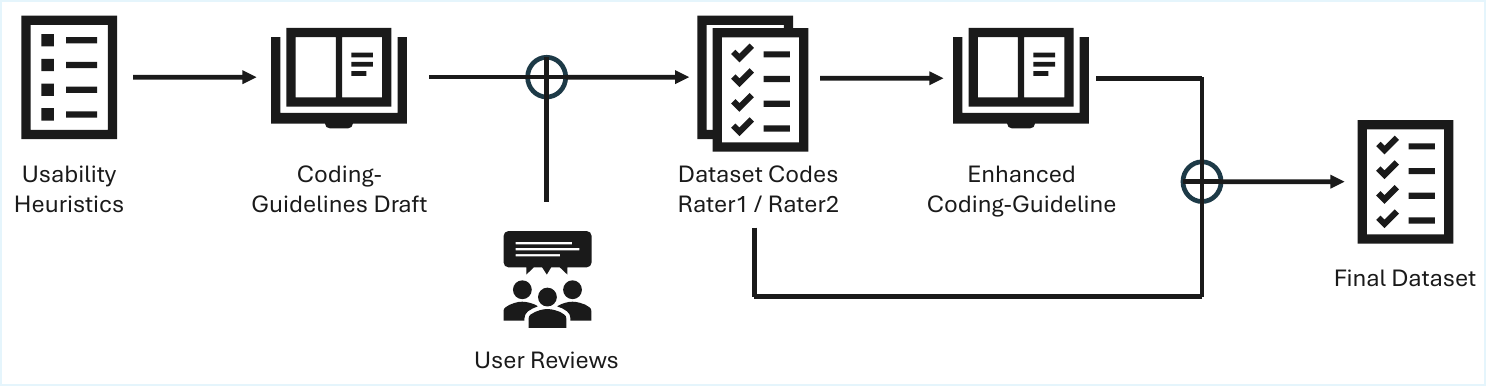}
    \caption{Data collection process for building a fully labeled dataset.}
    \label{fig:dc-process}
\end{figure}
\noindent
The labeling of the user reviews was performed by two raters, both with multiple coding experience in context of usability and explainability, which is why we believe that the results were not influenced by the quality of the raters.  Individual user review were rated in the binary format of \textit{true} and \textit{false}. We decided to code a user review with \textit{true}, if at least one usability-relevant statement is included in the review, otherwise we rate it as \textit{false}. For example, the review “I find the app rather cumbersome; some things could have been made easier” would be rated as \textit{true} because it relates to the heuristic “Flexibility and Efficiency of Use.” On the other hand, the review “Google Calendar works perfectly with aCalendar” would be rated \textit{false} because it does not directly address usability, but rather compatibility.

This first draft of coding guidelines was used by two raters in a pre-coding round. The pre-coding only looked at the first 20 user reviews of the \textit{BlitzerDE} app. We noted some differences in the understanding of the guidelines between raters. These issues were resolved by adding examples to the heuristics to the draft guidelines. After the draft guidelines were established, both raters fully labeled the dataset of 300 user reviews. To finalize the coding guidelines, the raters compared their results to each other. The comparison showed further issues with the draft guidelines. These are e.g. (a) bug-related reviews, (b) the handling of usability-problems which stem from outside of the apps context, e.g. Android, and (c) misleading user reviews. The coding guidelines were updated according to these issues and further examples were provided. With these enhanced coding guidelines, we conducted a second round of coding and calculated the interrater-reliability. To fulfill purpose (1), the last differences in the coding of both raters have been finalized into a single rating. In the finalization of our coding process, we discussed each user review with differing codes. We fulfill purpose (2) of the data collection process by comparing the results of the LLM with our final dataset and apply the following metrics: the interrater-reliability which treats the LLM as a human rater and allows direct comparisons to human performance, and the F-Score which is a common metric for binary classification performance and gives insights over the absolute ability of the LLM and the chosen prompt to label usability-relevant user reviews.
\newline\newline
\noindent
\textbf{Second Phase:} The second phase of our data analysis answers research questions RQ1 and RQ2. For this, the LLM is used to label the full dataset of 300 user reviews. To answer RQ1, the LLM is evaluated according to the ground-truth established in the data collection in terms of (M1) precision, (M2) accuracy, and (M3) interrater-reliability. Metrics (M1) and (M2) allow us to investigate if the LLM generally is able to distinguish usability-related reviews in a random sample, whereas metric (M3) gives insights about the LLMs reliability for identifying usability-related reviews when compared to human usability-experts.

\begin{figure}[h]
    \centering
    \includegraphics[width=0.9\linewidth]{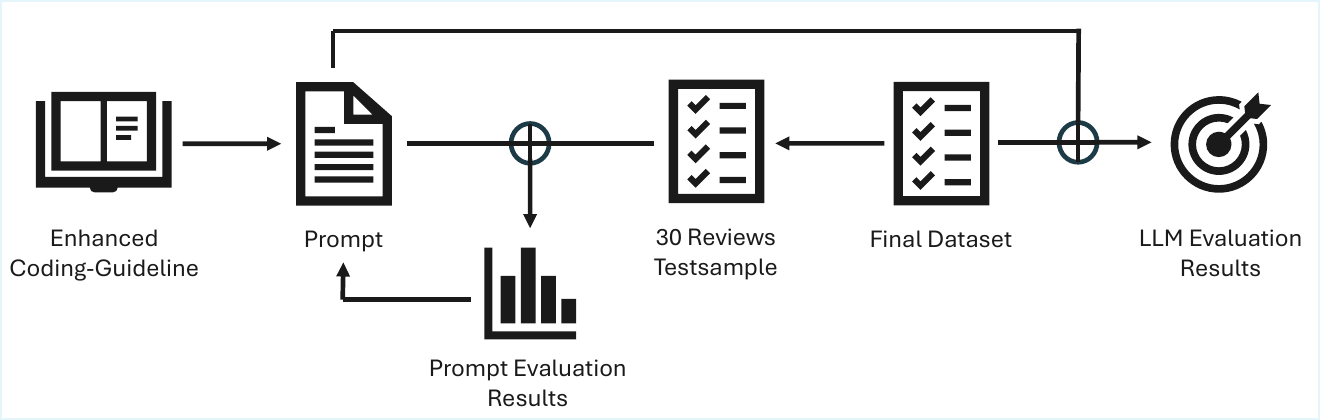}
    \caption{Data analysis process with the final dataset and an initial prompt as input to the iterativ prompt evaluation process, which results in a new and optimized prompt that can be fed back into the process to further refine it.}
    \label{fig:dc-process}
\end{figure}

\noindent
The process consists of two parts (i) generating and evaluating a prompt based on the enhanced coding guidelines from the data collection, and (ii) evaluating the LLM based on an specifically tailored prompt when tasked to label the final dataset from the data collection process. The specifically tailored prompt was generated through an iterative process to achieve a good performance. Based on the enhanced coding guidelines, an initial prompt was designed. To create the initial prompt, we used chain-of-thought prompting to give the LLM a clear list of instructions on how to work with the data. However, this first version solely served the purpose of being a template for further iterations. To evaluate prompt-iterations, the LLM was given the current iteration of the prompt and a subset of the fully labeled dataset. This test sample consists of 30 (10\%) randomly selected user reviews with 10 reviews from each app. For this paper, three iterations were conducted with the last iteration resulting in the final prompt. While evolving the prompt, we decided against using few-shot prompting. Instead, we added notes to the prompt that more clearly specified our goals, while being general enough to not cause overfitting inside the prompt.

After three iterations of adjusting the specifications inside the prompt, the final prompt and the full dataset were then used to evaluate the LLM according to the defined metrics (M1), (M2), and (M3).

\subsubsection{Data Availability Statement}

To ensure the transparency and verifiability of our research, we provide the following data in our supplementary material \cite{wellhausen_2026}: Firstly, we provide the Python script used to collect the user reviews from the Google Play-Store. We also provide the coding guidelines, as well as the full dataset containing all the tables with user reviews and their ratings in terms of usability-relevant aspects. Lastly, we provide all prompts that were used to detect usability aspects in user reviews with an LLM.
\section{Results}

\subsection{First Phase: Manually Labeled User Reviews}
\label{sec:first-phase}

Since the coding process involved two coders and the data was labeled on a nominal scale, we report the interrater reliability using Cohen’s Kappa $\kappa$ \cite{cohen60kappa}.
According to Landis and Koch \cite{landis1977measurement} we achieved a substantial agreement for \textit{BlitzerDE} (Cohen's kappa $\kappa$ = 0.76). \textit{Lidl Plus} also achieved a substantial agreement (Cohen's kappa $\kappa$ = 0.70) and \textit{FL Studio} also achieved a substantial agreement (Cohen's kappa $\kappa$ = 0.61) in our final coding round.

\begin{table}[h]
    \centering
    \caption{Interrater Agreement (Cohen's Kappa $\kappa$) for the manual detection.}
    {\renewcommand{\arraystretch}{1.2}
    \begin{tabular}{|l|l|l|l}
    \hline
        App & Cohen's Kappa $\kappa$ & Agreement according to Landis et al. \cite{landis1977measurement} \\
        \hline\hline
        BlitzerDE & 0.76 & Substantial \\
        Lidl Plus & 0.70 & Substantial \\
        FL Studio & 0.61 & Substantial \\
    \hline
    \end{tabular}
    }
    \label{tab:app_stats_irr_human}
\end{table}
\noindent
After resolving all conflicts, from the 300 user reviews, the two raters were able to extract 148 (49.3\%) user reviews in total which involved at least one usability aspect using the predefined coding guidelines. The final dataset contained 56 usability-relevant user reviews from \textit{BlitzerDE}, 54 user reviews from \textit{Lidl Plus}, and 38 reviews from \textit{FL Studio}.

\subsection{Second Phase: Automatic Labeling of User Reviews}

We chose to use gpt-4.1 by OpenAI to evaluate the user reviews as this was the most recent model at the time of conducting the evaluation and had a well documented API. This section presents (1) the results of the prompt generation which was tested on a randomized sample of 30 user reviews from the full dataset, and (2) the final LLM evaluation results of a full-scale test on the dataset. 

\begin{table}[h]
    \centering
    \caption{Results of the three prompt iterations on the Testsample.}
    \resizebox{\textwidth}{!}{%
    {\renewcommand{\arraystretch}{1.2}
    \begin{tabular}{|l||c|c|c||c|c|c||c|c|c|}
    \hline
        Iteration & \multicolumn{3}{|c||}{BlitzerDE (n=10)} & \multicolumn{3}{c||}{Lidl Plus (n=10)} & \multicolumn{3}{c|}{FL Studio (n=10)} \\ \cline{2-10}
        ~ & Precision & Recall & $F_1$-Score & Precision & Recall & $F_1$-Score & Precision & Recall & $F_1$-Score \\
        \hline
        Iteration 1             & 0.33 & 1.0 & 0.5 &    0.67 & 1.0 & 0.8 &      0.3 & 1.0 & 0.46 \\
        Iteration 2             & 0.375 & 1.0 & 0.55 &  0.75 & 1.0 & 0.86 &     0.33 & 1.0 & 0.5 \\
        \textbf{Iteration 3}    & 0.67 & 0.67 & 0.67 &  0.83 & 0.83 & 0.83 &    0.75 & 1.0 & 0.86 \\
    \hline
    \end{tabular}
    }
    }
    \label{tab:llm_pretest_scores}
\end{table}

\noindent
The initial prompt has been revised two times and has been evaluted three times at all stages. For our method of improving the initial prompt iteratively, we report the values of the Precision, Recall, and $F_1$-Score in Table~\ref{tab:llm_pretest_scores}. The values for each of the three apps are calculated separately.
\newline
\noindent
Our final evaluation on the full dataset with the gpt-4.1 resulted in the LLM identifying 173 (58\%) user reviews as usability-relevant, with 61 user reviews from \textit{BlitzerDE}, 68 reviews from \textit{Lidl Plus} and 44 user reviews from \textit{FL Studio}. Table~\ref{tab:llm_fulltest_scores} shows the Precision, Recall, and $F_1$-Score for this evaluation. In total, the LLM correctly labeled 239 (79.6\%) user reviews.

\begin{table}[h]
    \centering
    \caption{Results of the final evaluation on the full dataset.}
    \resizebox{\textwidth}{!}{%
    {\renewcommand{\arraystretch}{1.2}
    \begin{tabular}{|c|c|c||c|c|c||c|c|c|}
    \hline
        \multicolumn{3}{|c||}{BlitzerDE (n=100)} & \multicolumn{3}{c||}{Lidl Plus (n=100)} & \multicolumn{3}{c|}{FL Studio (n=100)} \\ \cline{1-9}
        Precision & Recall & $F_1$-Score & Precision & Recall & $F_1$-Score & Precision & Recall & $F_1$-Score \\
        \hline
        0.79 & 0.86 & 0.82 &    0.74 & 0.93 & 0.82 &      0.73 & 0.84 & 0.78 \\
    \hline
    \end{tabular}
    }
    }
    \label{tab:llm_fulltest_scores}
\end{table}
\noindent
For the interrater reliability with the LLM, we used the final coding result table of the two raters with all resolved conflicts and the coding result table generated by the LLM using the final prompt. Since this coding process also involved two coders, the combined codes by the human raters from the third coding round and the codes determined by the LLM, and the data also was labeled on a nominal scale, we report the interrater agreement using Cohen’s Kappa $\kappa$ \cite{cohen60kappa}. According to Landis and Koch \cite{landis1977measurement} the human codings combined with the LLM codings achieved a fair agreement for \textit{BlitzerDE} (Cohen's kappa $\kappa$ = 0.34). \textit{Lidl Plus} achieved a moderate agreement (Cohen's kappa $\kappa$ = 0.55) and \textit{FL Studio}  achieved a substantial agreement (Cohen's kappa $\kappa$ = 0.63).

\begin{figure}[h]
    \centering
    \includegraphics[width=0.9\linewidth]{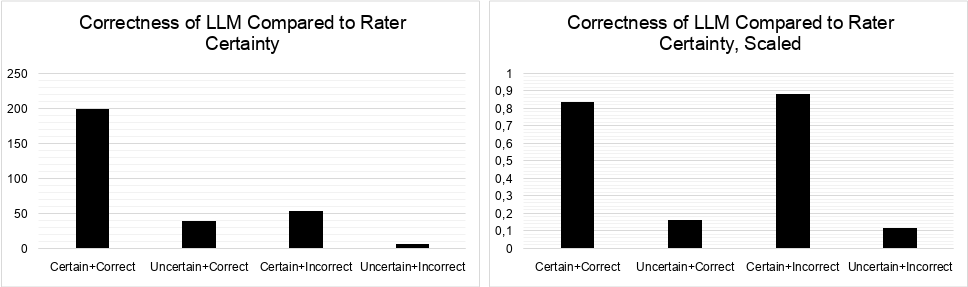}
    \caption{Distribution of the cross-examination of certainty of the raters vs. the correctness of the LLM's evaluation. Left side: Absolute values. Right side: Relatively values, scaled by 239 (cases of LLM being correct) and 61 (cases of LLM being incorrect).}
    \label{fig:llmcorrectness}
\end{figure}

\begin{table}[h]
    \centering
    {\renewcommand{\arraystretch}{1.2}
    \caption{Interrater Agreement (Cohen's Kappa $\kappa$) for the detection with an LLM and the final coding results of the third iteration. The $\kappa$ is calculated between the labels in the final dataset and the labels of the LLM.}
    \begin{tabular}{|l|l|l|l}
    \hline
        App & Cohen's Kappa $\kappa$ & Agreement according to Landis et al. \cite{landis1977measurement} \\
        \hline\hline
        BlitzerDE & 0.34 &  Fair \\
        Lidl Plus & 0.55 &  Moderate \\
        FL Studio & 0.63 &  Substantial \\
    \hline
    \end{tabular}
    }
    \label{tab:app_stats_irr_llm}
\end{table}
\noindent
Further, we report the observed values for when the two raters were certain in their final rating compared to the LLM labeling correctly according to the final rating. Figure~\ref{fig:llmcorrectness} shows the distribution of the four possible combinations for the accumulated values of the three apps. A rating is considered certain if both raters agreed on the final rating. The LLM labeling is considered correct if it equals the final rating in the fully labeled dataset. The $\chi^2$-Test results on these values in a $p$-value of 0.35. The observed values are therefore not statistically significant at $p < 0.05$.

\section{Discussion}

\subsection{Answering the Research Questions}

\paragraph{RQ1: Is the performance of an LLM, when prompted with a specifically tailored prompt to identify usability-related user reviews, comparable to the performance of a human coder?}

As can be seen in Table~\ref{tab:llm_fulltest_scores}, the LLM has achieved high values for the Recall measurements, which indicates a high correspondence with our findings of usability-relevant user reviews. Although not perfect, this shows that a significant portion of usability-relevant user reviews can be identified by an LLM with a prompt that is specifically tailored over multiple iterations. When combined with the Precision score, which indicates that the LLM's identified usability relevant user reviews were correct assessments, the LLM shows to be more permissive with its labeling than the human raters.

\paragraph{RQ2: What is the reliability of an LLM for identifying usability-related user reviews in a random dataset of user generated app reviews in case of specifically tailored prompts?}

Table~\ref{tab:app_stats_irr_llm} reports a mixed reliability when compared to the human raters. For BlitzerDE and Lidl Plus, the human raters had a significantly better agreement than the LLM with the fully labeled dataset. However, the LLM performed slightly better for the FL Studio partition of the dataset. With the LLM achieving Substantial and Moderate agreement for FL Studio and Lidl Plus respectively, a prompt with just three iterations already achieves reliable results. However, taking Figure~\ref{fig:llmcorrectness} into perspective, the LLM's correctness in labeling the user reviews does not correlate with the raters being certain in their rating. The LLM therefore doesn't reliably reproduce the mental model of the human raters and produces errors in different and unexpected places. This issue may be alleviated by improving the prompt in such a way, that the thought process of an usability-expert can be recreated by the LLM.

\subsection{Threats to Validity}

We discuss the threats to the validity of this work in accordance with Wohlin et al. \cite{wohlin2012experimentation}:

\paragraph{Construct Validity.}

Online user reviews are generally carried by emotions and are therefore easily influenced by both positive and negative emotions. To minimize this influence on the fully labeled dataset, we excluded to broadly formulated statements from being labeled as usability. We only focused on statements that directly contained a reference to one of the heuristics. Negative experiences with software are also more likely to be reported than good or neutral experiences. As negative experiences are commonly linked to bugs or unexpected behavior, we excluded bug-related statements from being labeled as usability-relevant, even though these could be classified under the heuristic \textit{Error Prevention}.

\paragraph{Internal Validity.}

Users may report problems as a review to an app, that actually stems from other components of the system, like the operating system or hardware issues. To mitigate this, we only included statements that are linked to the apps usability. Statements that were clearly related to other components of the system and only affected the apps usability as a side effect, were excluded from being labeled usability-relevant. Updates may introduce new bugs or features to an app. This can lead to a change in usability or a wave of both positive feedback, praising the new feature as well as negative feedback, criticizing the new feature or reporting bugs. To mitigate this selection bias, we chose to use the sorting by \textit{most relevance} for collecting user reviews, as the other options \textit{newest} and \textit{ratings} would introduce a selection bias for either the aforementioned problem as well as only retrieving the best or worst feedback. Another important factor is that the prompt to evaluate the user reviews was given to the LLM in english, but the user reviews are in german. As we did no direct comparison with a german prompt, mixing both english and german language for the input to the LLM, side-effects are generally possible. However, LLMs and specifically GPT-4.1 achieve high benchmarking scores in specifically multilingual benchmarking tests, such as the \textit{Multilingual MMLU}~\cite{introgpt41}.

\paragraph{Conclusion Validity.}

The conclusions drawn from the labeled dataset are heavily influenced by the quality of the raters. As there were only two raters, a significant shift in the results could be seen if one rater had significantly more or less familiarity with usability. We believe that the results were not influenced by the quality of the raters, as we designed the guidelines in a collaborative way that included detailed discussions of both the usability heuristics and when to apply them, as well as possible edge-case scenarios. This is supported by our substantial agreement in the final round of coding.

\paragraph{External Validity.}

Our results are not directly generalizable to the entirety of apps and LLMs. This is due to both the limited amount of apps we analyzed as well as the limited amount of LLMs we experimented on. As can be seen in our results, even in our small sample of apps, LLM performance may vary drastically. It is unclear to us if the observed effects are amplified when looking at a larger frame or eventually even out. For this work, we took apps with vastly different usage scenarios to gather a diverse set of user reviews to increase the generalization as much as possible.

\section{Conclusion and Future Work}

In this paper, we examined if using LLMs together with CrowdRE approaches can be an effective method for user-centric requirements engineering. We conclude that LLMs, although not perfect, should be considered as a viable option in requirements engineering for labeling unstructured data like user reviews. Our data shows that the reliability between the human raters and the LLM is too low as for the LLM to replace human raters yet. We still argue that the resource overhead (i.e. training) of using more traditional Deep-Learning or Machine-Learning approaches isn't practical for most requirements engineers, when an LLM can achieve acceptable results with significantly less overhead. We found that generating a tailored prompt is one of the defining aspects for a successful labeling task. In our results, the LLM tends to be slightly more permissive, which we do not interpret as a problem. In practice, we consider gathering a limited amount of irrelevant user review as better than missing a relevant user review.

This work serves as a precursor to further research into the LLM integration into the CrowdRE framework. Our results provide a fully labeled dataset which can be used as a performance baseline. In future work, we aim to increase the LLMs reliability with more sophisticated prompt engineering methodologies, such as few-shot prompting. We also want to compare the traditional methods of Deep-Learning and Machine-Learning with LLMs and optimized prompts. This in turn will give more detailed answers on the question of when to use which technology. It is also possible to take a more detailed look at different large-language models. As LLMs are gaining more attention, different companies train new models with different capabilities, that may be leveraged in CrowdRE.

\section*{Decleration on Generative AI}

During the preparation of this work, the authors used OpenAI's GPT-4.1 in order to label user reviews, which were then further analyzed. The authors did not use text produced by generative AI for this work.

\bibliography{sample-ceur}

@inproceedings{hedegaard,
    author = {Hedegaard, Steffen and Simonsen, Jakob Grue},
    title = {Extracting usability and user experience information from online user reviews},
    year = {2013},
    isbn = {9781450318990},
    publisher = {Association for Computing Machinery},
    address = {New York, NY, USA},
    url = {https://doi.org/10.1145/2470654.2481286},
    doi = {10.1145/2470654.2481286},
    booktitle = {Proceedings of the SIGCHI Conference on Human Factors in Computing Systems},
    pages = {2089–2098},
    numpages = {10},
    location = {Paris, France},
    series = {CHI '13}
}

@INPROCEEDINGS{bakiu,
    author={Bakiu, Elsa and Guzman, Emitza},
    booktitle={2017 IEEE 25th International Requirements Engineering Conference Workshops (REW)}, 
    title={Which Feature is Unusable? Detecting Usability and User Experience Issues from User Reviews}, 
    year={2017},
    volume={},
    number={},
    pages={182-187},
    doi={10.1109/REW.2017.76}
}

@book{nielsen_book,
    author = {Nielsen, Jakob},
    title = {Usability Engineering},
    year = {1994},
    isbn = {9780080520292},
    publisher = {Morgan Kaufmann Publishers Inc.},
    address = {San Francisco, CA, USA}
}

@misc{ten_ue_heuristics,
    title={10 usability heuristics for user interface design},
    url={https://www.nngroup.com/articles/ten-usability-heuristics/},
    journal={Nielsen Norman Group},
    author={Nielsen, Jakob},
    year={1994},
    note={last accessed: 01/12/2026}
}

@misc{introgpt41,
    title={Introducing GPT‑4.1 in the API},
    url={https://openai.com/index/gpt-4-1/},
    journal={OpenAI Research},
    author={OpenAI},
    year={2025},
    note={last accessed: 02/17/2026}
}

@book{wohlin2012experimentation,
  title={Experimentation in software engineering},
  author={Wohlin, Claes and Runeson, Per and H{\"o}st, Martin and Ohlsson, Magnus C and Regnell, Bj{\"o}rn and Wessl{\'e}n, Anders and others},
  volume={236},
  year={2012},
  publisher={Springer}
}

@article{iso20249241,
  title={{ISO 9241-110:2020 Ergonomics of human-system interaction — Part 110: Interaction principles}},
  author={ISO},
  journal={International Standards Organisation},
  year={2024}
}

@InProceedings{usabilitydefinition,
    author="Bevan, Nigel
    and Carter, James
    and Harker, Susan",
    editor="Kurosu, Masaaki",
    title="ISO 9241-11 Revised: What Have We Learnt About Usability Since 1998?",
    booktitle="Human-Computer Interaction: Design and Evaluation",
    year="2015",
    publisher="Springer International Publishing",
    address="Cham",
    pages="143--151",
    isbn="978-3-319-20901-2"
}

@article{QUINONES201789,
    title = {How to develop usability heuristics: A systematic literature review},
    journal = {Computer Standards \& Interfaces},
    volume = {53},
    pages = {89-122},
    year = {2017},
    issn = {0920-5489},
    doi = {https://doi.org/10.1016/j.csi.2017.03.009},
    url = {https://www.sciencedirect.com/science/article/pii/S0920548917301058},
    author = {Daniela Quiñones and Cristian Rusu},
}

@misc{white2023promptpatterncatalogenhance,
      title={A Prompt Pattern Catalog to Enhance Prompt Engineering with ChatGPT}, 
      author={Jules White and Quchen Fu and Sam Hays and Michael Sandborn and Carlos Olea and Henry Gilbert and Ashraf Elnashar and Jesse Spencer-Smith and Douglas C. Schmidt},
      year={2023},
      eprint={2302.11382},
      archivePrefix={arXiv},
      primaryClass={cs.SE},
      url={https://arxiv.org/abs/2302.11382}, 
}

@article{prompt,
    author = {Liu, Pengfei and Yuan, Weizhe and Fu, Jinlan and Jiang, Zhengbao and Hayashi, Hiroaki and Neubig, Graham},
    title = {Pre-train, Prompt, and Predict: A Systematic Survey of Prompting Methods in Natural Language Processing},
    year = {2023},
    issue_date = {September 2023},
    publisher = {Association for Computing Machinery},
    address = {New York, NY, USA},
    volume = {55},
    number = {9},
    issn = {0360-0300},
    url = {https://doi.org/10.1145/3560815},
    doi = {10.1145/3560815},
    month = jan,
    articleno = {195},
    numpages = {35}
}

@misc{sahoo2025systematicsurveypromptengineering,
      title={A Systematic Survey of Prompt Engineering in Large Language Models: Techniques and Applications}, 
      author={Pranab Sahoo and Ayush Kumar Singh and Sriparna Saha and Vinija Jain and Samrat Mondal and Aman Chadha},
      year={2025},
      eprint={2402.07927},
      archivePrefix={arXiv},
      primaryClass={cs.AI},
      url={https://arxiv.org/abs/2402.07927}, 
}

@misc{schulhoff2025promptreportsystematicsurvey,
      title={The Prompt Report: A Systematic Survey of Prompt Engineering Techniques},
      year={2025},
      eprint={2406.06608},
      archivePrefix={arXiv},
      primaryClass={cs.CL},
      url={https://arxiv.org/abs/2406.06608}, 
}

@inproceedings{Wei,
    author = {Wei, Jason and Wang, Xuezhi and Schuurmans, Dale and Bosma, Maarten and ichter, brian and Xia, Fei and Chi, Ed and Le, Quoc V and Zhou, Denny},
    booktitle = {Advances in Neural Information Processing Systems},
    editor = {S. Koyejo and S. Mohamed and A. Agarwal and D. Belgrave and K. Cho and A. Oh},
    pages = {24824--24837},
    publisher = {Curran Associates, Inc.},
    title = {Chain-of-Thought Prompting Elicits Reasoning in Large Language Models},
    url = {https://proceedings.neurips.cc/paper_files/paper/2022/file/9d5609613524ecf4f15af0f7b31abca4-Paper-Conference.pdf},
    volume = {35},
    year = {2022}
}

@inproceedings{nielsen,
    author = {Nielsen, Jakob},
    title = {Enhancing the explanatory power of usability heuristics},
    year = {1994},
    isbn = {0897916506},
    publisher = {Association for Computing Machinery},
    address = {New York, NY, USA},
    url = {https://doi.org/10.1145/191666.191729},
    doi = {10.1145/191666.191729},
    booktitle = {Proceedings of the SIGCHI Conference on Human Factors in Computing Systems},
    pages = {152–158},
    numpages = {7},
    location = {Boston, Massachusetts, USA},
    series = {CHI '94}
}

@article{landis1977measurement,
    title = {The Measurement of Observer Agreement for Categorical Data},
    author = {J. Richard Landis and Gary G. Koch},
    ISSN = {0006341X, 15410420},
    doi = {https://doi.org/10.2307/2529310}                                  ,
    journal = {Biometrics},
    number = {1},
    pages = {159--174},
    publisher = {[Wiley, International Biometric Society]},
    volume = {33},
    year = {1977}
}

@article{cohen60kappa,
    author = {Cohen, Jacob},
    year = {1960},
    title = {A Coefficient of Agreement for Nominal Scales},
    pages = {37--46},
    volume = {20},
    number = {1},
    issn = {0013-1644},
    journal = {Educational and Psychological Measurement},
    doi = {10.1177/001316446002000104}
}

@phdthesis{crowdre,
    title = "Crowd-Based Requirements Engineering",
    author = "Groen, \{Eduard Christiaan\}",
    year = "2025",
    month = sep,
    day = "22",
    doi = "10.33540/3091",
    language = "English",
    isbn = "978-90-393-7904-2",
    series = "SIKS Dissertation Series",
    publisher = "Utrecht University",
    number = "34",
    type = "Doctoral thesis 2 (Research NOT UU / Graduation UU)",
    school = "Universiteit Utrecht",
}

@article{nlp4re,
    author = {Zhao, Liping and Alhoshan, Waad and Ferrari, Alessio and Letsholo, Keletso J. and Ajagbe, Muideen A. and Chioasca, Erol-Valeriu and Batista-Navarro, Riza T.},
    title = {Natural Language Processing for Requirements Engineering: A Systematic Mapping Study},
    year = {2021},
    issue_date = {April 2022},
    publisher = {Association for Computing Machinery},
    address = {New York, NY, USA},
    volume = {54},
    number = {3},
    issn = {0360-0300},
    url = {https://doi.org/10.1145/3444689},
    doi = {10.1145/3444689},
    journal = {ACM Comput. Surv.},
    month = apr,
    articleno = {55},
    numpages = {41}
}

@misc{llm4re,
    title={Large Language Models (LLMs) for Requirements Engineering (RE): A Systematic Literature Review}, 
    author={Mohammad Amin Zadenoori and Jacek Dąbrowski and Waad Alhoshan and Liping Zhao and Alessio Ferrari},
    year={2025},
    eprint={2509.11446},
    archivePrefix={arXiv},
    primaryClass={cs.SE},
    url={https://arxiv.org/abs/2509.11446}, 
}

@inproceedings{unterbusch,
    author = {Unterbusch, Max and Sadeghi, Mersedeh and Fischbach, Jannik and Obaidi, Martin and Vogelsang, Andreas},
    year = {2023},
    month = {09},
    pages = {102-111},
    title = {Explanation Needs in App Reviews: Taxonomy and Automated Detection},
    doi = {10.1109/REW57809.2023.00024}
}

@inproceedings{intelligent_agents,
    author={Dąbrowski, Jacek and Cai, Wanling and Bennaceur, Amel and Nuseibeh, Bashar and Alrimawi, Faeq},
    booktitle={2025 IEEE 33rd International Requirements Engineering Conference (RE)}, 
    title={Intelligent Agents for Requirements Engineering: Use, Feasibility and Evaluation}, 
    year={2025},
    volume={},
    number={},
    pages={535-543},
    doi={10.1109/RE63999.2025.00064}
}

@INPROCEEDINGS{finetuning,
    author={Wei, Fusheng and Keeling, Robert and Huber-Fliflet, Nathaniel and Zhang, Jianping and Dabrowski, Adam and Yang, Jingchao and Mao, Qiang and Qin, Han},
    booktitle={2023 IEEE International Conference on Big Data (BigData)}, 
    title={Empirical Study of LLM Fine-Tuning for Text Classification in Legal Document Review}, 
    year={2023},
    volume={},
    number={},
    pages={2786-2792},
    doi={10.1109/BigData59044.2023.10386911}
}

@inproceedings{nielsen1994enhancing,
  title={Enhancing the explanatory power of usability heuristics},
  author={Nielsen, Jakob},
  booktitle={Proceedings of the SIGCHI conference on Human Factors in Computing Systems},
  pages={152--158},
  year={1994}
}

@inproceedings{SoftwareTypes,
author = {Forward, Andrew and Lethbridge, Timothy C.},
title = {A taxonomy of software types to facilitate search and evidence-based software engineering},
year = {2008},
isbn = {9781450378826},
publisher = {Association for Computing Machinery},
address = {New York, NY, USA},
url = {https://doi.org/10.1145/1463788.1463807},
doi = {10.1145/1463788.1463807},
booktitle = {Proceedings of the 2008 Conference of the Center for Advanced Studies on Collaborative Research: Meeting of Minds},
articleno = {14},
numpages = {13},
location = {Ontario, Canada},
series = {CASCON '08}
}

@misc{wellhausen_2026, title={Supplementary Material to User Reviews as a Source for Usability Requirements}, url={https://figshare.com/collections/Supplementary_Material_to_User_Reviews_as_a_Source_for_Usability_Requirements/8256262/2}, DOI={10.6084/m9.figshare.c.8256262.v2}, abstractNote={

This collection contains the supplementary material to our work User Reviews as a Source for Usability Requirements. It contains the fully labeled dataset, the latest iteration of our prompt, and the script used to data-mine the user reviews.}, publisher={figshare}, author={Wellhausen, Cedric}, year={2026}, month={Jan} }

\end{document}